# Three-qubit pure-state canonical forms for Perfect Teleportation

Xin-Wei Zha*, Hai-Yang Song

Department of Applied Mathematics and Physics, Xi'an Institute of Posts and Telecommunications,

Xi'an, 710061 Shaanxi, China

Recently, Agrawal and Pati [Phys. Rev. A **74**, 062320 (2006)] have given a class of W-states that can be used for perfect teleportation. Here, two canonical forms of perfect quantum channel are presented by transformation operator and the GHZ state and the W state are special case of those two canonical forms of perfect quantum channel. Furthermore, the orthogonal complete measurement bases are given.

## 1. Introduction

Since the first create of quantum teleportation protocol by Bennett [1], research on Quantum teleportation has been attracting much attention both theoretically and experimentally in recent years due to its important applications in quantum calculation and quantum communication. For example, already there have been several experimental implementations [2-3] of teleportation and several other schemes of quantum teleportation have been presented [4-9].

The original protocol of Bennett et al [1] involves teleportation of an unknown qubit using an EPR pair by sending two bits of classical information from Alice to Bob. A natural question is that in addition to the existing two particles entangled states whether one can exploit other multi particle entangled states for quantum teleportation. That one cans teleport an unknown qubit using three-particle GHZ state [10] and four-particle GHZ state [11] was shown to be possible. Dür et al [12] pointed out that the states of three qubits could be entangled in two inequivalent ways: GHZ-class and W-class. And that W-states cannot be converted to GHZ states under stochastic local operation and classical communication (SLOCC). Many quantum teleportation protocol using W-state have been studied [13,14]. But it was shown that one needs to do non-local operation to recover the unknown

* Corresponding Author. Tel: +86-29-88166094. *E-mail address:* zhxw@xiyou.edu.cn (Xin-Wei Zha)

state. It was also shown that if one uses W-state then the teleportation protocol works with non-unit fidelity, i.e., it is not perfect. Recently, Agrawal and Pati [15] show that by performing von Neumann projection on the three-particles and by sending two bits of classical information one can teleport an unknown qubit perfectly using a class of W-states. But they did not give the general forms three-qubit state being suitable for that. In this paper, we give the two canonical forms of three-qubit pure state for perfect teleportation.

## 2. Perfect teleportation of three-qubit states quantum channel

Let us consider a situation where Alice has particles '1' and '2' and Bob has the particle '3'. Alice also wishes to teleport the unknown state of a particle 'a' in the following unknown state

$$|\chi\rangle_a = (\alpha|0\rangle + \beta|1\rangle)_a \tag{1}$$

Suppose that Alice and Bob share a three-qubit entangled state $|\varphi\rangle_{123}$ given by

$$|\varphi\rangle_{123} = a_0|000\rangle + a_1|001\rangle + a_2|010\rangle + a_3|011\rangle + a_4|100\rangle + a_5|101\rangle + a_6|110\rangle + a_7|111\rangle \tag{2}$$

The system state of four particles is

$$|\psi\rangle_s = |\varphi\rangle_{123} \otimes |\chi\rangle_a \tag{3}$$

In accordance with the principle of superposition and transformation operator [16, 17], the wave function of $|\psi\rangle_s$ can be represented in the form of a series

$$|\psi\rangle_s = |\varphi\rangle_{123} \otimes |\chi\rangle_a = \frac{1}{2}\sum_{i=1}^{8} \varphi_{12a}^i \hat{\sigma}_3^i |\chi\rangle_3 \tag{4}$$

Where $\varphi_{12a}^i$ are the orthogonal complete basis, and

$$|\chi\rangle_3 = (\alpha|0\rangle + \beta|1\rangle)_3 \tag{5}$$

the $\sigma_3^i$ is called the "transformation operator". In order to realize the teleportation, Alice can make a measurement on the three particles 'a12' and convey her results to Bob via classical communication. Then the resulting state of Bob's particles 3 will be states $|\psi\rangle_3^i = \sigma_3^i |\chi\rangle_3$.

Now the question is: what are the states that Alice and Bob can share that can be used as a quantum resource to transfer the state of particle 'a' to particle '3' using usual teleportation protocol.

From Eq. (4) can known, If the $\sigma_3^i (i=1,2,3,4)$ is unitary operator and the $\sigma_3^i (i=5,6,7,8)$ is zero operator, the $|\psi\rangle_s$ can be represented in the form of a series

$$|\psi\rangle_s = |\varphi\rangle_{123} \otimes |\chi\rangle_a = \frac{1}{2}\sum_{i=1}^{4}|\varphi_{12a}^i\rangle U_3^i|\chi\rangle_3 \qquad (6)$$

then Alice informs Bob the measurement outcomes via a classical channel. By outcomes received, Bob can determine the state of particles 3 exactly by unitary transformation operators $(U_3^i)^{-1}$. The unknown particle entangled state can be teleported perfectly, and both the successful possibilities and the fidelities reach unit.

## 3. The two canonical forms of perfect quantum teleportation states

It is well known, the canonical forms of three-qubit pure state [18]

$$|\psi\rangle = \kappa_0 e^{i\theta}|000\rangle + \kappa_1|100\rangle + \kappa_2|010\rangle + \kappa_3|100\rangle + \kappa_4|111\rangle \qquad (7)$$

Analogous, there are two forms of perfect entanglement channel to satisfy Eq. (6).

### 3.1 The first canonical form of perfect quantum entanglement channel is

$$|\psi\rangle_{123} = a|000\rangle + be^{i\delta}|010\rangle + \sqrt{\frac{1}{2} - (a^2+b^2)}e^{i\lambda}|100\rangle + \frac{\sqrt{2}}{2}e^{i\gamma}|111\rangle)_{123} \qquad (8a)$$

where a, b is a real number and $\delta$, $\lambda$ and $\gamma$ are phases.

Alice can make a von-Neumann type measurement using the states $|\varphi^i\rangle_{12a}$ $i=1,2,\cdots 8$, which are given by

$$|\varphi^1\rangle_{12a} = a|000\rangle + be^{i\delta}|010\rangle + \sqrt{\frac{1}{2} - (a^2+b^2)}|100\rangle + \frac{\sqrt{2}}{2}e^{i\gamma}|111\rangle)_{123}$$

$$|\varphi^2\rangle_{12a} = a|000\rangle + be^{i\delta}|010\rangle + \sqrt{\frac{1}{2} - (a^2+b^2)}e^{i\lambda}|100\rangle - \frac{\sqrt{2}}{2}e^{i\gamma}|111\rangle)_{123}$$

$$|\varphi^3\rangle_{12a} = a|001\rangle + be^{i\delta}|011\rangle + \sqrt{\frac{1}{2} - (a^2+b^2)}e^{i\lambda}|101\rangle + \frac{\sqrt{2}}{2}e^{i\gamma}|110\rangle)_{123}$$

$$|\varphi^4\rangle_{12a} = a|001\rangle + be^{i\delta}|011\rangle + \sqrt{\frac{1}{2}-(a^2+b^2)}e^{i\lambda}|101\rangle - \frac{\sqrt{2}}{2}e^{i\gamma}|110\rangle)_{123}$$

$$|\varphi^5\rangle_{12a} = c_0|000\rangle + c_2e^{i\delta}|010\rangle + c_4e^{i\lambda}|100\rangle$$

$$|\varphi^6\rangle_{12a} = d_0|000\rangle + d_2e^{i\delta}|010\rangle + d_4e^{i\lambda}|100\rangle$$

$$|\varphi^7\rangle_{12a} = c_1|001\rangle + c_3e^{i\delta}|011\rangle + c_5e^{i\lambda}|101\rangle)_{12a}$$

$$|\varphi^8\rangle_{12a} = d_1|001\rangle + d_3e^{i\delta}|011\rangle + d_5e^{i\lambda}|101\rangle)_{12a}$$

For the above eight states are mutually orthogonal, the coefficients must satisfy

$$c_0 a + c_2 b + c_4 \sqrt{\frac{1}{2}-(a^2+b^2)} = 0,$$

$$d_0 a + d_2 b + d_4 \sqrt{\frac{1}{2}-(a^2+b^2)} = 0,$$

$$c_0 d_0 + c_2 d_2 + c_4 d_4 = 0 \tag{8b}$$

Then we have

$$U_3^1 = \begin{pmatrix} 1 & 0 \\ 0 & 1 \end{pmatrix}, U_3^2 = \begin{pmatrix} 1 & 0 \\ 0 & -1 \end{pmatrix}, U_3^3 = \begin{pmatrix} 0 & 1 \\ 1 & 0 \end{pmatrix}, U_3^4 = \begin{pmatrix} 0 & -1 \\ 1 & 0 \end{pmatrix} \tag{8c}$$

Therefore this class of states can be used perfect quantum teleportation.

### 3.2 The second canonical form of perfect quantum entanglement channel is

$$|\psi\rangle_{123} = a|001\rangle + be^{i\delta}|010\rangle + \sqrt{\frac{1}{2}-b^2}e^{i\lambda}|100\rangle + \sqrt{\frac{1}{2}-a^2}e^{i\gamma}|111\rangle)_{123} \tag{9a}$$

Alice can make a von-Neumann type measurement using the states $|\varphi^i\rangle_{12a}$ $i=1,2,\cdots 8$, which are given by

$$|\varphi^1\rangle_{12a} = a|001\rangle + be^{i\delta}|010\rangle + \sqrt{\frac{1}{2}-b^2}e^{i\lambda}|100\rangle + \sqrt{\frac{1}{2}-a^2}e^{i\gamma}|111\rangle)_{123}$$

$$\left|\varphi^{2}\right\rangle_{12a} = -a\left|001\right\rangle + be^{i\delta}\left|010\right\rangle + \sqrt{\frac{1}{2}-b^{2}}\,e^{i\lambda}\left|100\right\rangle - \sqrt{\frac{1}{2}-a^{2}}\,e^{i\gamma}\left|111\right\rangle)_{123}$$

$$\left|\varphi^{3}\right\rangle_{12a} = a\left|000\right\rangle + be^{i\delta}\left|011\right\rangle + \sqrt{\frac{1}{2}-b^{2}}\,e^{i\lambda}\left|101\right\rangle + \sqrt{\frac{1}{2}-a^{2}}\,e^{i\gamma}\left|110\right\rangle)_{123}$$

$$\left|\varphi^{3}\right\rangle_{12a} = -a\left|000\right\rangle + be^{i\delta}\left|011\right\rangle + \sqrt{\frac{1}{2}-b^{2}}\,e^{i\lambda}\left|101\right\rangle - \sqrt{\frac{1}{2}-a^{2}}\,e^{i\gamma}\left|110\right\rangle)_{123}$$

$$\left|\varphi^{5}\right\rangle_{12a} = \sqrt{2}(\sqrt{\frac{1}{2}-a^{2}}\left|001\right\rangle - ae^{i\gamma}\left|111\right\rangle)_{12a}$$

$$\left|\varphi^{6}\right\rangle_{12a} = \sqrt{2}(\sqrt{\frac{1}{2}-b^{2}}\,e^{i\delta}\left|010\right\rangle - be^{i\lambda}\left|100\right\rangle)_{12a}$$

$$\left|\varphi^{7}\right\rangle_{12a} = \sqrt{2}(\sqrt{\frac{1}{2}-a^{2}}\left|000\right\rangle - ae^{i\gamma}\left|110\right\rangle)_{12a},$$

$$\left|\varphi^{8}\right\rangle_{12a} = \sqrt{2}(\sqrt{\frac{1}{2}-b^{2}}\,e^{i\delta}\left|011\right\rangle - be^{i\lambda}\left|101\right\rangle)_{12a} \tag{9b}$$

Then, we have

$$U_{3}^{1} = \begin{pmatrix} 1 & 0 \\ 0 & 1 \end{pmatrix}, U_{3}^{2} = \begin{pmatrix} 1 & 0 \\ 0 & -1 \end{pmatrix}, U_{3}^{3} = \begin{pmatrix} 0 & 1 \\ 1 & 0 \end{pmatrix}, U_{3}^{4} = \begin{pmatrix} 0 & -1 \\ 1 & 0 \end{pmatrix} \tag{9c}$$

Obviously, in (8a), if $b = 0, a = \frac{\sqrt{2}}{2}$, then

$$\left|\psi\right\rangle_{123} = \frac{\sqrt{2}}{2}\left|000\right\rangle + \frac{\sqrt{2}}{2}e^{i\gamma}\left|111\right\rangle)_{123} \tag{10}$$

It is just that of GHZ state [10].

At the same time,

$$\left|\varphi^{1}\right\rangle_{12a} = \frac{\sqrt{2}}{2}\left|000\right\rangle + \frac{\sqrt{2}}{2}e^{i\gamma}\left|111\right\rangle)_{12a}$$

$$\left|\varphi^{2}\right\rangle_{12a} = \frac{\sqrt{2}}{2}\left|000\right\rangle - \frac{\sqrt{2}}{2}e^{i\gamma}\left|111\right\rangle)_{123}$$

$$\left|\varphi^3\right\rangle_{12a} = \frac{\sqrt{2}}{2}|001\rangle + \frac{\sqrt{2}}{2}e^{i\gamma}|110\rangle)_{123}$$

$$\left|\varphi^4\right\rangle_{12a} = \frac{\sqrt{2}}{2}|001\rangle - \frac{\sqrt{2}}{2}e^{i\gamma}|110\rangle)_{12a}$$

$$\left|\varphi^5\right\rangle_{12a} = |010\rangle$$

$$\left|\varphi^6\right\rangle_{12a} = |011\rangle$$

$$\left|\varphi^7\right\rangle_{12a} = |100\rangle)_{12a}$$

$$\left|\varphi^8\right\rangle_{12a} = |101\rangle)_{12a} \tag{11}$$

Therefore, $\left|\varphi^1\right\rangle_{12a}, \left|\varphi^2\right\rangle_{12a}, \left|\varphi^3\right\rangle_{12a}, \left|\varphi^4\right\rangle_{12a}$ is just those of $\left|\psi_1^\pm\right\rangle_{a12}, \left|\psi_2^\pm\right\rangle_{a12}$ [15], where

$$\left|\psi_1^\pm\right\rangle = \frac{1}{\sqrt{2}}(|000\rangle \pm |111\rangle)_{a12}$$

$$\left|\psi_2^\pm\right\rangle = \frac{1}{\sqrt{2}}(|100\rangle \pm |011\rangle).$$

From (9) (if $a = \frac{\sqrt{2}}{2}$), it can be seen that

$$|\psi\rangle_{123} = \frac{\sqrt{2}}{2}|001\rangle + be^{i\delta}|010\rangle + \sqrt{\frac{1}{2}-b^2}\,e^{i\lambda}|100\rangle \tag{12}$$

Then from (9a), we have

$$\left|\varphi^1\right\rangle_{12a} = \frac{\sqrt{2}}{2}|001\rangle + be^{i\delta}|010\rangle + \sqrt{\frac{1}{2}-b^2}\,e^{i\lambda}|100\rangle$$

$$\left|\varphi^2\right\rangle_{12a} = -\frac{\sqrt{2}}{2}|001\rangle + be^{i\delta}|010\rangle + \sqrt{\frac{1}{2}-b^2}\,e^{i\lambda}|100\rangle$$

$$\left|\varphi^3\right\rangle_{12a} = \frac{\sqrt{2}}{2}|000\rangle + be^{i\delta}|011\rangle + \sqrt{\frac{1}{2}-b^2}\,e^{i\lambda}|101\rangle$$

$$|\varphi^4\rangle_{12a} = -\frac{\sqrt{2}}{2}|000\rangle + be^{i\delta}|011\rangle + \sqrt{\frac{1}{2}-b^2}\,e^{i\lambda}|101\rangle$$

$$|\varphi^5\rangle_{12a} = |111\rangle_{12a}$$

$$|\varphi^6\rangle_{12a} = \sqrt{2}(\sqrt{\frac{1}{2}-b^2}\,e^{i\delta}|010\rangle - be^{i\lambda}|100\rangle)_{12a}$$

$$|\varphi^7\rangle_{12a} = |110\rangle_{12a},$$

$$|\varphi^8\rangle_{12a} = \sqrt{2}(\sqrt{\frac{1}{2}-b^2}\,e^{i\delta}|011\rangle - be^{i\lambda}|101\rangle)_{12a} \tag{13}$$

in (12), if $b=0$, then

$$|\psi\rangle_{123} = \frac{1}{\sqrt{2}}(|001\rangle + +|100\rangle)_{123} = \frac{1}{\sqrt{2}}|0\rangle_2(|01\rangle + +|10\rangle)_{13} \tag{14}$$

It is a Bell channel.

in (12), if $b=\frac{1}{2}$, then

$$|\psi\rangle_{123} = \frac{\sqrt{2}}{2}|001\rangle + \frac{1}{2}e^{i\delta}|010\rangle + \frac{1}{2}e^{i\lambda}|100\rangle \tag{15}$$

This state is $|W_1\rangle_{123}$ [15], and

$$|W_1\rangle_{123} = \frac{1}{2}(|100\rangle + |010\rangle + \sqrt{2}|001\rangle)_{123} \tag{16}$$

We can use either the state given in (8) or (9) for perfect quantum teleportation.

## 4. Conclusion

We have presented two canonical forms of three-qubit pure states quantum channel, which are used for perfect quantum teleportation. Two forms of quantum channel can not be converted from each other by local operations. But they can be converted by non-locality operations of $U_{12}$ [19]. As ref [15], it is easy to show that the reduced density matrix $\rho_3 = tr_{12}(|W_n\rangle\langle W_n|) = \frac{I}{2}$ for those perfect quantum channels. Obviously, the von Neumann entropy of $\rho_3$ is just one. That is there is one ebit of entanglement shared between Alice and Bob when they use those classes of states.

## Acknowledgements

This work is supported by Shaanxi Natural Science Foundation under Contract Nos. 2004A15 and Science Plan Foundation of office the Education Department of Shaanxi Province Contract Nos. 05JK288.